\documentclass[conference]{IEEEtran}
\IEEEoverridecommandlockouts

\usepackage{cite}
\usepackage[acronym]{glossaries}
\usepackage{algorithmic}
\usepackage{graphicx}
\usepackage{textcomp}
\usepackage[dvipsnames]{xcolor}
\usepackage{tikz}
\usepackage{pifont}
\newcommand{\cmark}{\ding{51}}%
\newcommand{\xmark}{\ding{55}}%

\usepackage{cleveref}
\def\BibTeX{{\rm B\kern-.05em{\sc i\kern-.025em b}\kern-.08em
    T\kern-.1667em\lower.7ex\hbox{E}\kern-.125emX}}

\newacronym{5g}{5G}{5th Generation}
\newacronym{6g}{6G}{6th Generation}
\newacronym{3gpp}{3GPP}{3rd Generation Partnership Project}
\newacronym{phy}{PHY}{Physical Layer}
\newacronym{ran}{RAN}{Radio-Access-Networks}
\newacronym{sdr}{SDR}{Software Defined Radio}
\newacronym{mimo}{MIMO}{Multiple Input Multiple Output}
\newacronym{mmse}{MMSE}{Minimum Mean Squared Error}
\newacronym{ofdm}{OFDM}{Orthogonal Frequency Division Multiplexing}
\newacronym{kpi}{KPI}{Key Performance Indicator}
\newacronym{mac}{MAC}{Multiply and Accumulate}
\newacronym{pynq}{PYNQ}{Python Productivity for Zynq}
\newacronym{hls}{HLS}{High Level Synthesis}
\newacronym{hdl}{HDL}{Hardware Description Language}
\newacronym{nr}{NR}{New Radio}
\newacronym{E2E}{E2E}{end-to-end}
\newacronym{bs}{BS}{basestation}
\newacronym{ue}{UE}{user equipment}
\newacronym{csi}{CSI}{channel state information}

\newacronym{mvm}{MVM}{Matrix Vector Multiplication}
\newacronym{ber}{BER}{Bit Error Rate}
\newacronym{mc}{MC}{Monte Carlo}
\newacronym{cots}{COTS}{Commercial Off The Shelf}
\newacronym{gpp}{GPP}{General Purpose Programmable}
\newacronym{raw}{RAW}{Read After Write}
\newacronym{wdotp}{wDotp}{widening-dotproduct}

\newacronym{isa}{ISA}{Instruction Set Architecture}
\newacronym{rtl}{RTL}{Register Transfer Level}
\newacronym{tlm}{TLM}{Transaction Level Modeling}
\newacronym{fpga}{FPGA}{Field Programmable Gate Array}
\newacronym{asic}{ASIC}{Application Specific Integrated Circuit}
\newacronym{sbt}{SBT}{Static Binary Translation}
\newacronym{ir}{IR}{Intermediate Representation}

\newacronym{fp}{FP}{Floating Point}
\newacronym{ipu}{IPU}{Integer Processing Unit}
\newacronym{fpu}{FPU}{Floating Point Unit}
\newacronym{dma}{DMA}{Direct Memory Access}
\newacronym{simd}{SIMD}{Single Instruction Multiple Data}
\newacronym{lsu}{LSU}{Load\&Store Unit}
\newacronym{dsp}{DSP}{Digital Signal Processing}
\newacronym{snr}{SNR}{Signal to Noise Ratio}
\newacronym{mips}{MIPS}{Million Instructions Per Second}
\newacronym{tti}{TTI}{Transmission Time Interval}
\newacronym{qam}{QAM}{Quadrature-Amplitude Modulation}

\newacronym{dut}{DUT}{Device Under Test}
\newacronym{awgn}{AWGN}{Additive White Gaussian Noise}
\usepackage{fontawesome}

\newcommand{\greencheck}{
  \tikz[baseline=-0.4ex] \node[draw=white, fill=ForestGreen, text=white, circle, inner sep=0.2mm]{\cmark};
}
\newcommand{\redcross}{
  \tikz[baseline=-0.6ex] \node[draw=white, fill=BrickRed, text=white, circle, inner sep=0.2mm]{\xmark};
}
\newcommand{\greenup}{\textcolor{ForestGreen}{\faArrowUp}}
\newcommand{\greendown}{\textcolor{ForestGreen}{\faArrowDown}}
\newcommand{\redup}{\textcolor{BrickRed}{\faArrowUp}}
\newcommand{\reddown}{\textcolor{BrickRed}{\faArrowDown}}

\makeatletter
\newcommand{\linebreakand}{%
  \end{@IEEEauthorhalign}
  \hfill\mbox{}\par
  \mbox{}\hfill\begin{@IEEEauthorhalign}
}

\newcommand\copyrighttext{\footnotesize \textcopyright This work was submitted to the IEEE for publication. Copyright may be transferred without notice, after which this version may no longer be accessible.}
\newcommand\copyrightnotice{%
    \begin{tikzpicture}[remember picture,overlay]
        \node[anchor=south,yshift=10pt] at (current page.south) {\fbox{\parbox{\dimexpr\textwidth-\fboxsep-\fboxrule\relax}{\copyrighttext}}};
    \end{tikzpicture}%
}

\begin{document}

\title{Fast End-to-End Simulation and Exploration of Many-RISCV-Core Baseband Transceivers for Software-Defined Radio-Access Networks
}

\author{
\IEEEauthorblockN{Marco Bertuletti}
\IEEEauthorblockA{
ETH Z\"{u}rich\\
Z\"{u}rich, Switzerland\\
mbertuletti@iis.ee.ethz.ch} \\
\and
\IEEEauthorblockN{Yichao Zhang}
\IEEEauthorblockA{
ETH Z\"{u}rich\\
Z\"{u}rich, Switzerland\\
yiczhang@iis.ee.ethz.ch}\\
\and
\IEEEauthorblockN{Mahdi Abdollahpour}
\IEEEauthorblockA{
Universit\`a di Bologna \\
Bologna, Italy \\
mahdi.abdollahpour@unibo.it}\\
\and
\IEEEauthorblockN{Samuel Riedel}
\IEEEauthorblockA{
ETH Z\"{u}rich\\
Z\"{u}rich, Switzerland\\
sriedel@iis.ee.ethz.ch}\\
\linebreakand
\IEEEauthorblockN{Alessandro Vanelli-Coralli}
\IEEEauthorblockA{ETH Z\"{u}rich\\
Z\"{u}rich, Switzerland\\
Universit\`a di Bologna \\
Bologna, Italy \\
avanelli@iis.ee.ethz.ch}
\and
\IEEEauthorblockN{Luca Benini}
\IEEEauthorblockA{ETH Z\"{u}rich\\
Z\"{u}rich, Switzerland\\
Universit\`a di Bologna \\
Bologna, Italy \\
lbenini@iis.ee.ethz.ch}
}



\maketitle

\begin{abstract}
The fast-rising demand for wireless bandwidth~\cite{Ericsson_MobilityReport_2024} requires rapid evolution of high-performance baseband processing infrastructure. Programmable many-core processors for software-defined radio (SDR) have emerged as high-performance baseband processing engines, offering the flexibility required to capture evolving wireless standards and technologies~\cite{Kamaleldin_manycore_2020, Zhang_TeraPool_2024, Rajagopal_EDGEQ_2021}. This trend must be supported by a design framework enabling functional validation and end-to-end performance analysis of SDR hardware within realistic radio environment models. We propose a static binary translation based simulator augmented with a fast, approximate timing model of the hardware and coupled to wireless channel models to simulate the most performance-critical physical layer functions implemented in software on a many (1024) RISC-V cores cluster customized for SDR. Our framework simulates the detection of a 5G OFDM-symbol on a server-class processor in 9.5s-3min, on a single thread, depending on the input MIMO size (three orders of magnitude faster than RTL simulation). The simulation is easily parallelized to 128 threads with 73-121$\times$ speedup compared to a single thread.
\end{abstract}


\copyrightnotice

\section{Introduction}

The worldwide demand for more wireless subscriptions (8.5 billion in 2023) and network data traffic (145EB in Q1-2024)~\cite{Ericsson_MobilityReport_2024} fostered the development of new \gls{ran} technologies. The Massive \gls{mimo} paradigm ratified by \gls{3gpp} Release-17~\cite{3GPP_TS_R17_description} gained momentum and will extend to more users and antennas in the upcoming \gls{6g} releases~\cite{Shen_6Gresearch_ACM_2022}. Alongside increased performance requirements, channel traffic characteristics will also change, resulting in a dynamic range from a few kBs for a voice call to GBs for high-quality multimedia~\cite{Wittig_modeling5G_2020}.

The flexibility required to address demanding worst-case specifications and wide operation ranges can be obtained by implementing the \gls{mimo} processing workload on programmable cores, an approach known as \gls{sdr}. Multi-core and many-core systems promise high performance on \gls{sdr} \gls{5g} workloads at the edge, implementing the full \gls{phy} on \glspl{bs}~\cite{Kamaleldin_manycore_2020, Zhang_TeraPool_2024, Rajagopal_EDGEQ_2021}. Additionally, the RISC-V \gls{isa} emerged as a powerful domain-specialization tool to fight back the energy-efficiency loss of instruction processors vs. custom hardware, enabling the co-design of efficiency-boosting processors features and domain-specific instruction extensions~\cite{Zhang_TeraPool_2024, Rajagopal_EDGEQ_2021}.

Evaluating the algorithmic (functional) performance of software running on \gls{sdr}-hardware and its execution performance under varying network traffic, spatial topology, and transmission channel conditions is critical to the design of future \glspl{bs}. To this purpose, a model of the \gls{sdr} receivers must be simulated as \gls{dut} in an \gls{E2E} transmission, matching the following requirements: deterministic behavior, flexibility to evolving standards, and awareness of timing to synchronize with other elements of the processing chain~\cite{Wittig_modeling5G_2020}. Additionally, a fast simulation framework is needed to evaluate the receiver performance via extensive numerical \gls{mc} simulation.

\begin{table}[h!]
\centering
\vspace{-1em}
\caption{Simulation methods for \gls{sdr} baseband hardware.}
\resizebox{\columnwidth}{!}{
\begin{tabular}{l l c c c c}
\hline
 & \textbf{Device} & \textbf{Freq.} & \textbf{Speed} & \textbf{Design Effort} & \textbf{Multi-Core}\\
\hline\hline
\cite{Rezgui_modelsim_2018, Park_verilog_2021} RTL                     & -             & -       & \reddown  & \greendown  & \redcross     \\\hline
\cite{Barreteau_systemC_2013} TLM                   & Intel Core2   & 3.00GHz & \greenup  & \redup      & \redcross     \\\hline
\cite{cheng_PYNQ_2023} FPGA                         & XCZU28DR      & 128MHz  & \greenup  & \greendown  & \redcross     \\\hline
\cite{Kamaleldin_manycore_2020} FPGA                & ZCU102        & 120MHz  & \greenup  & \redup      & \greencheck   \\\hline
\textbf{Ours SBT}                                   & AMD EPYC-7742 & 3.25GHz & \greenup  & \greendown  & \greencheck   \\\hline
\end{tabular}
}
\vspace{0.5em}
\label{tab:summary}
\end{table}

\Cref{tab:summary} summarizes different options to include \gls{sdr} hardware as \gls{dut} in an \gls{E2E} wireless transmission testbench. Event-based (e.g., QuestaSim~\cite{modelsim}) and compiled (e.g., Verilator~\cite{verilator}) \gls{rtl} simulators, require low effort to directly simulate \gls{hdl} code of baseband receivers~\cite{Rezgui_modelsim_2018, Park_verilog_2021, synopsys_keysight}. Unfortunately, they are too slow for \gls{mc} simulation of complex workloads.
\gls{tlm} in SystemC allows the creation of detailed virtual prototypes of entire digital systems~\cite{Barreteau_systemC_2013} and can also be used for the simulation of RF analog transceivers \cite{Miorandi_systemC_2018, Beichler_systemC_2021, Gailliard_systemC_2007}. However, \gls{tlm} simulations are limited in speed for the sequential execution of SystemC processes managed by a centralized kernel~\cite{Lonardi_systemC_2015} and require long compilation time, constraining design iterations.
\gls{fpga}~\cite{cheng_PYNQ_2023, Kamaleldin_manycore_2020} or \gls{asic} prototyping can not be used to implement a fast hardware-software co-design loop, because it requires a fully designed system and can be used only late in the design process. In the case of complex baseband many-core receivers, the prototyping on \gls{fpga} also requires considerable design effort. \gls{hls} tools, such as \gls{pynq}~\cite{cheng_PYNQ_2023}, are used to emulate small dataflows on \gls{fpga}, but the approach does not extend to many-core baseband processors with hundreds of processing elements. Moreover, implementing multi-core systems on \gls{fpga} must trade off limited hardware resources and operating frequencies, as clearly shown by the implementation of 36 RI5CY cores on Xilinx XCVU9, presented in \cite{Kamaleldin_manycore_2020}. 


This paper proposes a lightweight \gls{sbt} based framework to simulate \gls{sdr} many-core programmable transceivers based on the RISC-V \gls{isa}. Our simulator can be connected to channel models for extensive \gls{mc} analysis and is easily parallelized on \gls{cots} multi-core servers. 
The following contributions quantify these claims:
\begin{itemize}

\item 
We propose the open-source\footnote{https://github.com/pulp-platform/banshee} \gls{sbt}-based simulator Banshee~\cite{Banshee_Riedel_2021} for deterministic and instruction accurate emulation of many-core transceivers. The simulator is connected to wireless channel models~\cite{sionna} for \gls{E2E} \gls{mc} analysis of the receiver performance. As a challenging benchmark, we emulate the 1024 independent instruction streams of the open-source\footnote{https://github.com/pulp-platform/mempool} programmable TeraPool-\gls{sdr}~\cite{Zhang_TeraPool_2024} on a 128-cores AMD EPYC-7742 server.
\item 
We demonstrate fast design space exploration on the \gls{mmse} detection in different arithmetic precisions. A single thread of the 128-core AMD EPYC-7742 server simulates the detection of an \gls{ofdm} symbol in \textless~3min, depending on the \gls{mimo} size. A small 4x4 input only requires 9.5s, compared to 13h:44min for \gls{rtl} simulation of the same binary (Questasim-2022.3). Independent symbols are easily parallelized, with 73-121$\times$ speedup, compared to a single thread.
\item 
We show that augmenting the \gls{sbt}-based simulator with an approximate timing model of the hardware offers first-order estimates of the receiver processing time, with 30\% average error over the measured \gls{rtl} cycle-count for different \gls{mmse} implementations and \gls{mimo} sizes.
\end{itemize}

Running \gls{mmse} detection with a speedup over \gls{rtl} simulation exceeding 1000$\times$, the developed framework enables fast \gls{mc} analysis of programmable baseband transceivers.

\section{Many-Core SDR Architecture}

TeraPool~\cite{Zhang_TeraPool_2024}, the largest shared-memory many-core cluster for \gls{sdr} presented in the open literature, is a challenging simulation benchmark: emulating its 1024 independent instruction streams, diverse \gls{isa}, and large hierarchical interconnect is time-consuming in \gls{rtl} simulation and would require consistent modeling effort in \gls{tlm}. Moreover, TeraPool's open-source \gls{rtl} design enables comparisons with cycle-accurate simulation. 

\begin{figure}[ht]
  \centering
  \includegraphics[width=\linewidth]{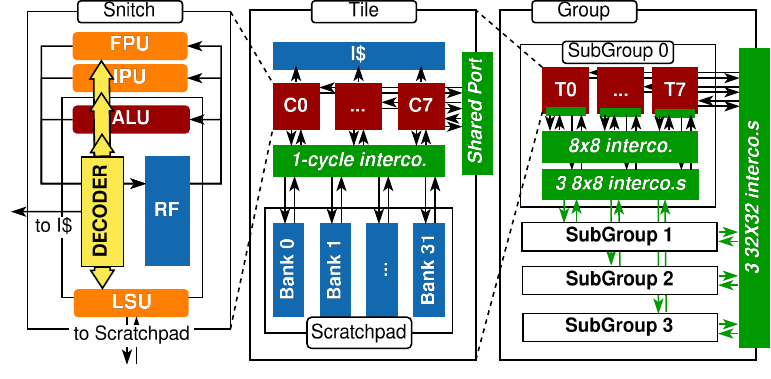}
  \caption{Snitch core, Tile, SubGroup, and Group of TeraPool.}
  \label{fig:archi}
  \vspace{-1em}
\end{figure}

The TeraPool architecture, illustrated in \Cref{fig:archi}, features \textit{Snitch} processing elements that support the RV32IMAF \gls{isa}. The single-stage core decodes instructions and executes the RV32I \gls{isa}, excluding load\&stores. Other instructions are offloaded to co-processing functional units, including a \gls{lsu}, an \gls{ipu} for integer multiplication and division, and a \gls{fpu} implementing the \textit{zfinx} and \textit{zhinx} extensions~\cite{RISCV_zfinx}. 

Snitch's functional units also implement custom integer and \gls{fp} RISC-V extensions for \gls{dsp}. The \textit{Xpulpimg} set includes post-increment load\&stores, \glspl{mac}, and \gls{simd} compute and data-shuffling operations on 8b-16b integer types. The \textit{Smallfloat} and \textit{Minifloat} sets~\cite{Tagliavini_smallfloats_2024, Bertaccini_minifloats_2024} include \gls{simd} operations on \gls{fp} 8b-16b types, \gls{wdotp} with 8b-16b operands and 16b-32b accumulators, and complex 16b \gls{fp} \gls{mac}.

The base of TeraPool's hierarchical design is the \textit{Tile}, where 8 cores share 32~KiB of scratchpad memory with 1-cycle access latency and 4~KiB of I\$. The cores access the scratchpad of other Tiles via shared ports to the cluster interconnects. One 8x8 crossbar connects 8 Tiles in the \textit{SubGroup} hierarchy. Three 8x8 crossbars connect Tiles of the SubGroup to Tiles of 3 other SubGroups in the \textit{Group} hierarchy. Three 32x32 crossbars connect the Tiles of a Group to Tiles in 3 other Groups, summing up to 128 Tiles and 4~MiB of scratchpad in a Cluster. Pipeline stages at hierarchy boundaries configure non-uniform memory access (less than 9 cycles without contentions). A cluster-level AXI interconnect and a \gls{dma} engine allow explicit data transfers from L2 memory.

\section{Co-simulation of DUT with TX-Channel}

We describe a wireless \gls{E2E} \gls{mimo} transmission and the developed framework for co-simulation of the transmission channel model and of the RISC-V \gls{dut}.

\subsection{End-to-end \gls{mimo} wireless transmission}

\gls{mimo} is a key enabling technology for 5G and beyond networks, where the \gls{bs} is equipped with $N_{RX}$ antenna elements, which receive signals from $N_{TX}$ \glspl{ue}. During uplink, the transmission bit sequence maps to a \gls{qam} constellation, and blocks of \gls{qam}-symbols are transmitted through the channel on $N_{SC}$ \gls{ofdm} subcarriers, namely an \gls{ofdm}-symbol~\cite{3GPP_TS_R17_description}. At the \gls{bs}, the detection algorithm uses \gls{csi} to estimate the transmitted \gls{qam}-symbols. The \gls{bs} processes a \gls{tti} with 14 \gls{ofdm}-symbols in \textless 1ms. The evolution from \gls{5g} to \gls{6g} will increase in the $N_{SC}$, $N_{TX}$, and $N_{RX}$, affecting the \glspl{kpi} of \gls{dsp} detection algorithms as well as their hardware implementation. 

We therefore identify \gls{mimo} detection as a reference application to prove the capabilities of our co-simulation approach. We implement the \gls{mmse} detector, which offers optimal performance for transmission through a \gls{awgn} channel, on the target \gls{sdr}-hardware.

\begin{figure}[ht]
  \centering
  \includegraphics[width=\linewidth]{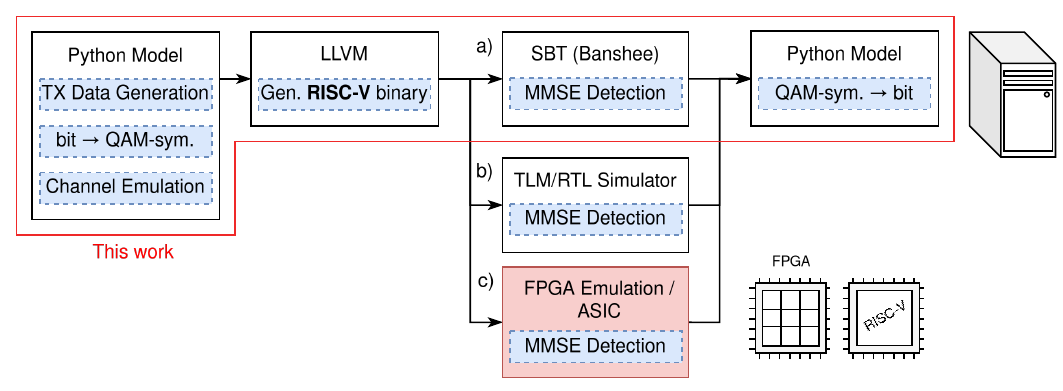}
  \caption{Different approaches to add RISC-V hardware in the loop of a telecommunication application: c) execution on prototype \gls{asic}, emulation on \gls{fpga}, b) cycle-accurate \gls{rtl}/\gls{tlm} simulation and a) simulation on Banshee.}
  \label{fig:loop}
  \vspace{-1em}
\end{figure}

\subsection{Simulation of the \gls{mimo} wireless transmission}

The proposed framework and flow are described in~\Cref{fig:loop}. A Python model of the \gls{mimo} transmission based on \cite{sionna} runs on the simulation host. The model includes bit-sequence transmission generation for different users, \gls{qam}-symbol mapping, and wireless channel emulation. It also implements a double-precision reference of the detection \gls{mmse} algorithm.

The \gls{sdr} hardware running \gls{mmse} processing is simulated on Banshee, an open-source LLVM-based binary translator for instruction-accurate emulation of multi-core systems. The simulation flow in Banshee is divided into two phases. During \textit{translation}, Banshee converts a RISC-V binary into LLVM \gls{ir} via \gls{sbt}. The generated \gls{ir} references to Banshee runtime functions to keep track of the emulated device state. LLVM optimizes and translates the \gls{ir} into host code. During \textit{emulation}, Banshee maps the translated code to multiple threads on the host platform for each hardware thread of the emulated RISC-V multi-core.

Besides offering an instruction-accurate simulation of the host and taking advantage of fully parallel execution, Banshee assigns a static latency to each instruction to estimate the program runtime. To model outstanding instruction execution, Banshee implements a scoreboard that keeps track of the \gls{raw} dependencies between instructions. This feature allows a fast software performance estimation by accounting for instruction stalls caused by long-latency instructions such as loads\&stores.

We modified Banshee to translate \textit{zfinx}, \textit{zhix}, \textit{Smallfloat}, and \textit{Minifloat} \gls{fp} extensions, which compute on operands in the integer register file, as specified in~\cite{RISCV_zfinx}. The latency of \gls{fp} instructions is also annotated in the simulator for first-order estimation on the emulated code runtime.

\begin{figure}[ht]
  \centering
  \includegraphics[width=\linewidth]{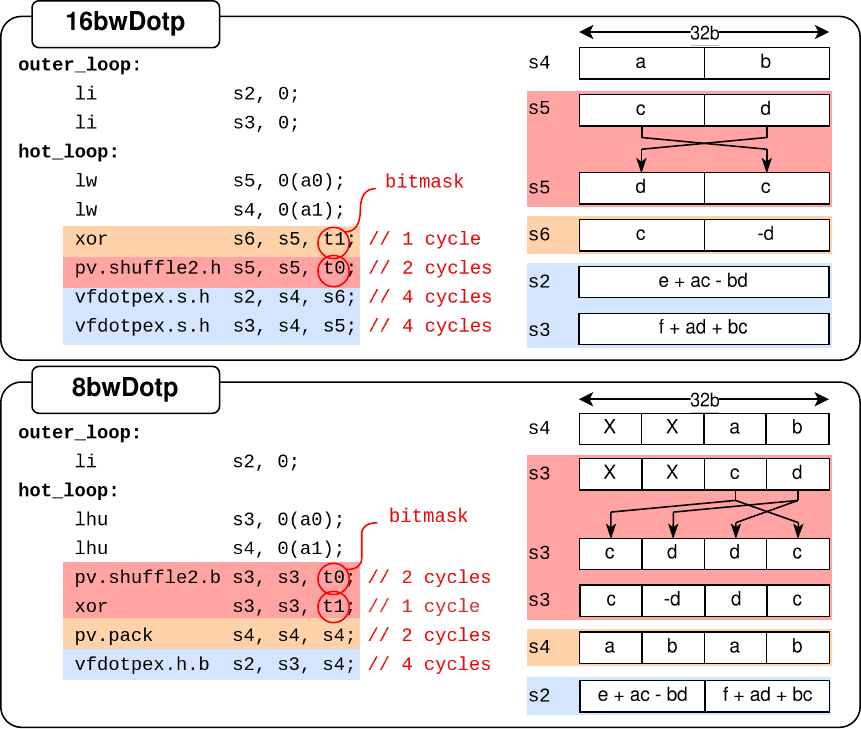}
  \caption{Execution of the complex \gls{mac} in different arithmetic precisions using \textit{Smallfloat} and \textit{Xpulpimg} instruction sets.}
  \label{fig:cmul}
  \vspace{-1.5em}
\end{figure}
\begin{figure}[ht]
  \centering
  \includegraphics[width=\linewidth]{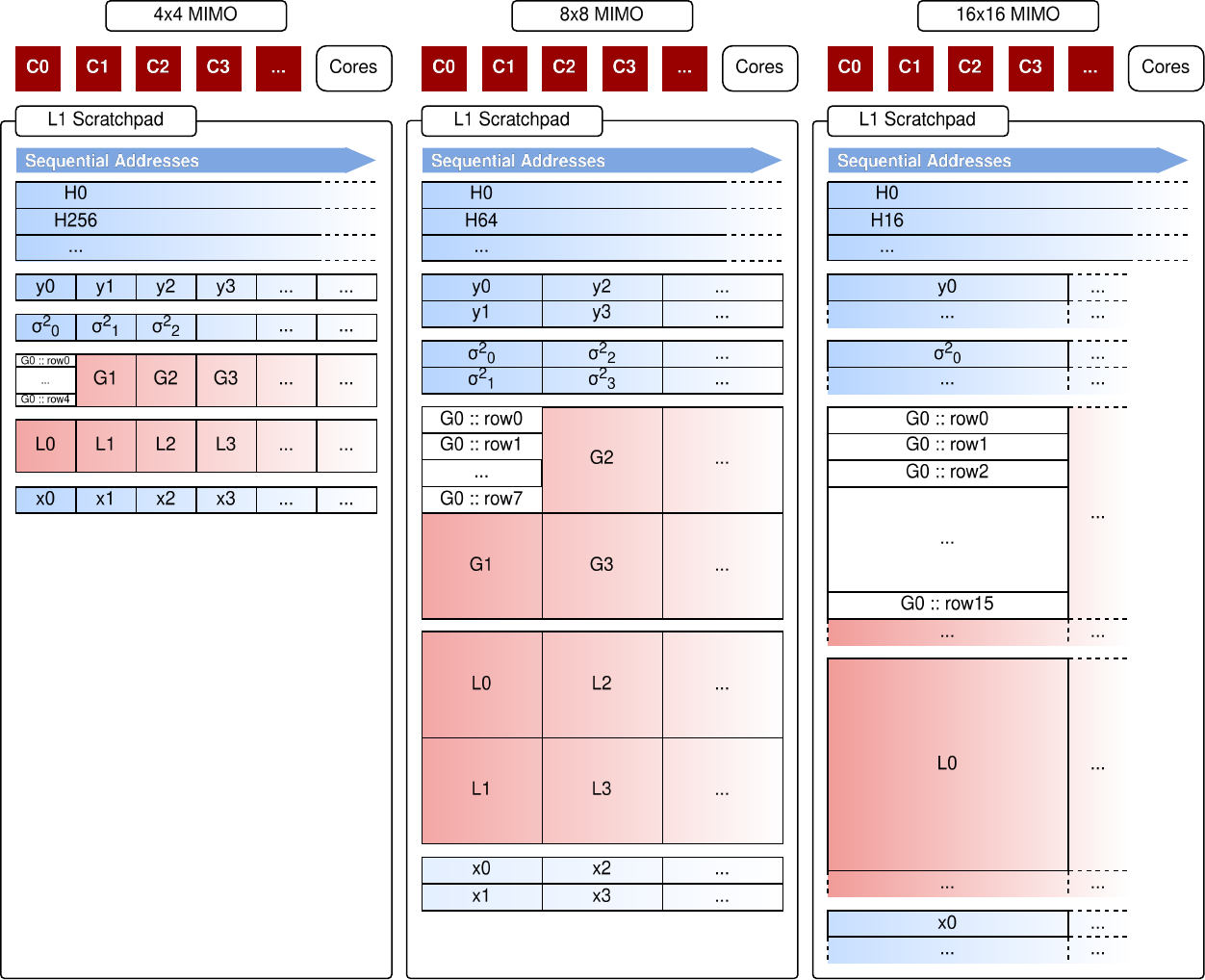}
  \caption{Implementation of parallel \gls{mimo}-\gls{mmse} on the TeraPool many-core cluster for different sizes of the \gls{mimo} problem.}
  \label{fig:mmse}
  \vspace{-1em}
\end{figure}

\section{Software-defined \gls{mmse} on TeraPool}

This section describes the \gls{sdr} implementation of a \gls{mmse} detector on TeraPool. Given $y$ the received signal, $\hat{H}$ and $\sigma$ an estimation of the channel and noise at the \gls{bs}, and $\hat{x}$ the detected signal, the analytical \gls{mmse} detector is:

\begin{equation}
    \hat{x}_i = (\hat{H_i}^H\hat{H_i} + \sigma_i^2I)^{-1}\hat{H_i}^Hy_i \quad \forall i \in [1, N_{SC}]
    \label{eq:mmse}
\end{equation}

The operator in \Cref{eq:mmse} requires a matrix inversion. Given $G = (\hat{H}^H\hat{H} + \sigma^2I)$ and $z = \hat{H}^Hy$, we decompose $G=L^HL$ in its lower and upper triangular components, using the Cholesky algorithm. The linear system $G\hat{x} = z$ is solved by inverting two lower triangular matrices $\hat{x}=L^{-1}\left((L^H)^{-1}z\right)$. This reduces the overall arithmetic complexity of the inversion, as described in \cite{Shahabuddin_MIMO_NorCAS_2020}. The \gls{mmse} detection then requires four operators: the Hermitian of the channel, the \gls{mvm} between the complex-conjugate of the channel and the received symbol, the Cholesky decomposition, and the solution of a triangular system. 

TeraPool has a \textit{fork-join} type programming model (common trait of many-core clusters). In the \textit{fork} phase of the program, each core executes an independent instruction stream in parallel. The \textit{join} phase consists of a synchronization barrier: cores wait for each other before executing the next parallel section. 

In an \gls{ofdm} transmission a different \gls{mmse} problem must be solved for each subcarrier of the transmission spectrum. In the parallel execution on TeraPool, the kernels of \gls{mmse} are implemented for a single Snitch core, and an independent \gls{mmse} problem is then distributed in parallel to each core. The single-core \gls{mmse} is implemented in different arithmetic precisions, as this is a key exploration parameter for trading off functional accuracy with execution speed and energy.

\subsubsection{16bHalf} uses the RISCV \textit{zhinx} set. We load 16b words. A complex \gls{mac} requires four \texttt{fmadd.h} instructions.
\subsubsection{16bwDotp} uses the \gls{wdotp} extension of the \textit{Smallfloat} set, with 32b accumulation. We load 32b words, and we use two \glspl{wdotp} and a \gls{simd} \texttt{shuffle} instruction from \textit{Xpulpimg} set to implement the complex \gls{mac}, as described in \Cref{fig:cmul}.
\subsubsection{16bCDotp} uses the complex dotproduct operation. The multiplication internal precision is 32b, and real-imaginary accumulators are in 16b, fitting a 32b word.
\subsubsection{8bQuarter} uses \textit{Smallfloat} 1b sign, 4b exponent, 2b mantissa \gls{fp} format. After the execution of the $\hat{H}^H\hat{H}$ matrix product and of the $\hat{H}^Hy$ \gls{mvm}, we cast the outputs to 16b to solve the $G\hat{x}=z$ linear system in higher numerical precision. 
\subsubsection{8bwDotp} uses \textit{Smallfloat} \gls{wdotp} for the matrix products and the \gls{mvm}, with accumulators in 16b. A complex \gls{mac} requires a \gls{wdotp} instruction and a \gls{simd} \texttt{shuffle} instruction from the \textit{Xpulpimg} set, as described in \Cref{fig:cmul}. The linear system is solved in 16b precision.

For the parallel execution, each core runs an independent \gls{mmse} problem (1024 in total). The operands are allocated in the shared memory of TeraPool to reduce the cores' contentions to shared interconnect resources~\cite{Bertuletti_PUSCH_2023}. In \Cref{fig:mmse}, we represent TeraPool cores and the corresponding input-outputs of independent \gls{mimo} problems. The elements of vectors in \Cref{eq:mmse} ($y$, $\sigma^2$, $\hat{x}$, and $\hat{H}$, flattened to a one-dimension array in row-major order) lay in consecutive addresses of L1. This matches the allocation of data in L2 and does not require relocating elements after explicit \gls{dma} transfers to L1. On the contrary, intermediate outputs of the computation $L$ matrix and $G$ matrix have rows on different memory rows. This keeps the data local to the processing cores. For example, in the 4x4 \gls{mimo}, each core sends requests to different banks, fetching $y$, $\sigma^2$, $\hat{x}$, and $G$, and in the 32x32 \gls{mimo} only 8 cores contend for the same scratchpad banks.

\begin{figure}[ht]
  \centering
  \includegraphics[width=\linewidth]{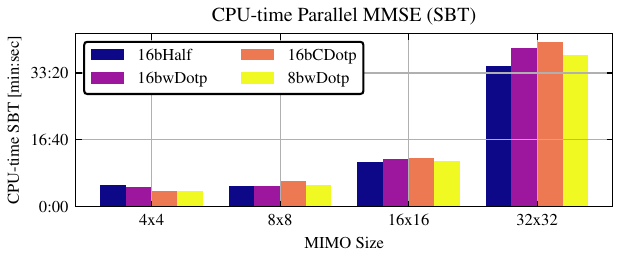}
  \includegraphics[width=\linewidth]{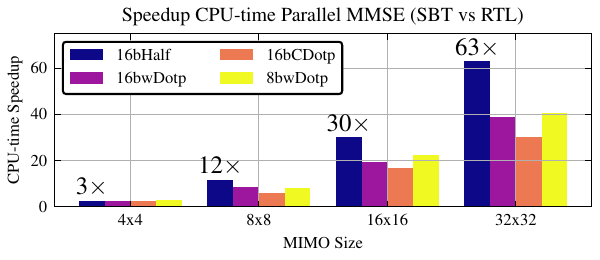}
  \caption{Parallel \gls{mmse} in different precisions and input sizes: speedup on 128-cores AMD EPYC-7742 of multi-thread Banshee simulation against single-thread \gls{rtl} simulation with QuestaSim-2022.3.}
  \label{fig:mmse_speedup}
  \vspace{-1em}
\end{figure}
\begin{figure}[ht]
  \centering
  \includegraphics[width=\linewidth]{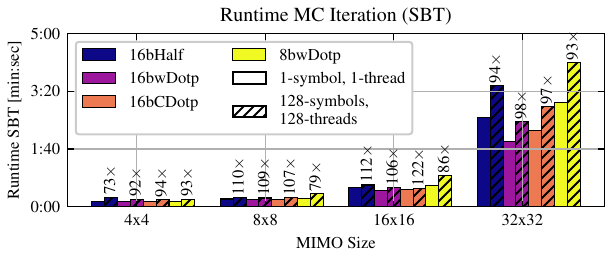}
  \caption{Batched execution on a Snitch of $N_{SC}=1638$ \gls{mmse} problems of an OFDM-symbol: single-thread runtime of Banshee simulation and multi-thread simulation of independent symbols with corresponding speedup, on a 128-cores AMD EPYC-7742 server.}
  \label{fig:mmse_mc}
\end{figure}

\section{BER \& Runtime Analysis}

This section explains how \gls{sbt} can be used for fast design-space exploration of \gls{mmse} in different arithmetic precisions. It presents Banshee runtime, then shows how to measure the \glspl{kpi} of software-defined \gls{mimo} applications.

\subsection{Simulator runtime performance}

We execute 1024 \gls{mmse} problems in parallel on 1024 TeraPool cores. We simulate a total of 1024 independent instruction streams. The experiment runs on a 128-CPUs \gls{cots} server processor at 3.25GHz (AMD EPYC-7742). Each CPU has 32MiB L3-cache and shares 1TB of system memory. We compare the total CPU-time of a multi-thread Banshee simulation with the CPU-time of a single-thread \gls{rtl} simulation (Questasim-2022.3 being intrinsically single-threaded). For multi-thread, we average the results of 10 independent runs to filter the effect of competing processes on the server. 
\Cref{fig:mmse_speedup} shows the runtime (top) and speedup (bottom) results:
the total CPU-time for a Banshee call amounts to average \textless~35min, while the average wall-clock runtime is \textless~2min:45s, over all the input sizes. Comparing CPU-times, Banshee simulation is up to 63$\times$ faster than single-thread QuestaSim-2022.3 \gls{rtl} simulation. Comparing wall-clock runtimes, Banshee multi-thread simulation is up to 1582$\times$ faster than single thread \gls{rtl} simulation.

To speed up the simulation, the \gls{mc} simulation of independent \gls{mmse} problems is batched on a single TeraPool core. We consider a \gls{nr} transmission in a 50MHz bandwidth, with $N_{SC}=1638$, 30kHz subcarrier spacing, and 0.5ms \gls{tti} duration.
We simulate a \gls{mc} iteration, consisting of $N_{SC}=1638$ \gls{mmse} problems running on a single Snitch, with a single AMD EPYC-7742 thread. The maximum speed for a single-threaded Banshee simulation of a single Snitch is $3.57$ \gls{mips}. The runtime in \Cref{fig:mmse_mc}, shows \textless~3min per \gls{mc} iteration, and down to 9.44s for an input 4x4 \gls{mimo}, compared to 13h:44min required by \gls{rtl} simulation of the same binary (5237$\times$ faster). 

Independent \gls{ofdm} symbols can be easily parallelized over the 128 cores of the server. \Cref{fig:mmse_mc} shows the runtime with 128 active threads and the speedup compared to a single thread. We measure 12.5s-4min:10s simulation runtime per 128 \gls{mc} iterations, depending on the \gls{mimo} size and arithmetic precision: 73-121$\times$ speedup compared to single-thread execution. 
The achieved simulation throughput is fully compatible with a fast HW-in-the-loop exploration of the \gls{E2E} \gls{5g} application under exam.

\subsection{\gls{mmse} cycle-count on TeraPool}

\begin{figure}[ht]
  \centering
  \includegraphics[width=\linewidth]{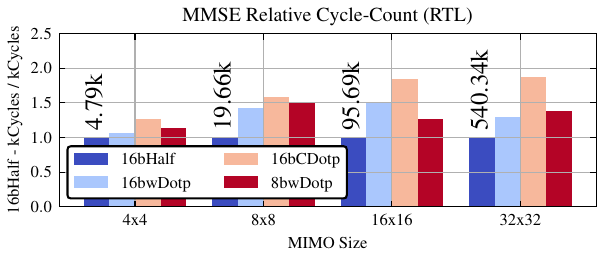}
  \includegraphics[width=\linewidth]{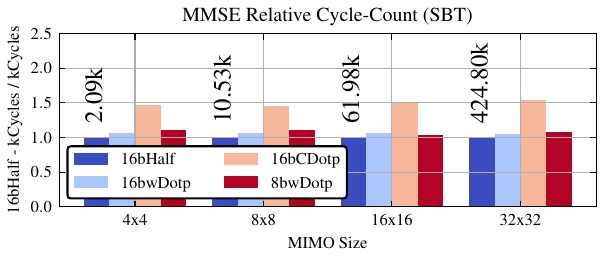}
  \includegraphics[width=\linewidth]{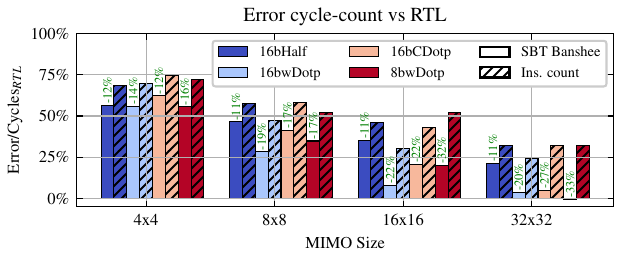}
  \caption{Parallel implementation of the \gls{mmse} in different precisions for various sizes of \gls{mimo}. Error between the cycles measured by \gls{rtl} simulation with QuestaSim-2022.3 and estimated by Banshee or by counting instructions.}
  \label{fig:parmmse_cycles}
  \vspace{-1em}
\end{figure}

\Cref{fig:parmmse_cycles} reports the cycle-count of the complete parallel \textit{16bHalf} \gls{mmse} application, measured by \gls{rtl} simulation in QuestaSim-2022.3. The cycle-accurate \gls{rtl} simulation gives precise information on the speedup of lower arithmetic precisions compared to \textit{16bHalf}. \glspl{simd} reduce the total number of instructions issued. In particular, for a 32x32 input, the ratio between instructions issues in the \textit{16bHalf} compared to \textit{16bwDotp}, \textit{16bCDotp}, and \textit{8bwDotp} corresponds to 1.15$\times$, 1.84$\times$, and 1.37$\times$. These values are slightly smaller than the measured speedup (1.29$\times$, 1.86$\times$, 1.38$\times$): requiring fewer loads, compared to \textit{16bHalf}, \gls{simd} instructions also reduce the stalls from contentions in the shared interconnect. 

\begin{figure}[ht]
  \centering
  \includegraphics[width=\linewidth]{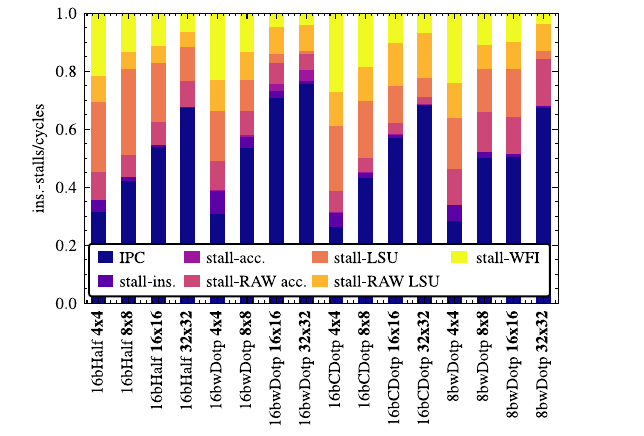}
  \caption{Breakdown of instructions and architectural stalls over the cycle-count. Result of cycle-accurate experiments in QuestaSim-2022.3.}
  \label{fig:mmse_ipc}
  \vspace{-1em}
\end{figure}

The breakdown of instructions and stalls over the cycle count for different arithmetic precisions and input problem sizes is in \Cref{fig:mmse_ipc}. We measure few \textit{stall-ins.} for I\$ refill and \textit{stall-acc.}, occurring when the pipelines of Snitch functional units are full. Loops are unrolled to minimize \gls{raw} stalls, with increasing benefits at higher problem sizes. The amount of stalls due to contentions in the interconnect (\textit{stall-LSU}) is the highest for the low arithmetic intensity \textit{16bHalf} application. Finally, \textit{stall-WFI} refers to the time spent by cores idling at the synchronization during the fork-join \gls{mmse} application.

\Cref{fig:parmmse_cycles} reports a comparison between the cycle count measured by \gls{rtl} simulation, the cycle count estimated by Banshee, and a rough estimation based on the instruction count. Privileging multi-thread fast simulation of parallel programs on the host, Banshee omits priorities between loads\&stores targeting the same memory bank or Tiles' shared-ports to interconnects, as well as priorities between atomic memory operations, used to implement synchronization barriers. Modeling these effects would require expensive synchronization between hardware threads or computationally heavy statistical models. Our approximate solution, oriented to simulation speed, is conservative in assigning statically to all the transactions the largest memory access latency without contentions (9 cycles).

On average, we obtain 30\% error compared to \gls{rtl} simulation cycle-count. The largest errors measured (20-62\%) correspond to the \textit{16bHalf} implementation, which loads\&stores separately the real-imaginary parts of the operands (twice the number of memory operations, compared to other kernels). Small \gls{mimo} problems with large synchronization overhead are more affected by the permissive latency modeling of atomics. Nevertheless, in this unfavorable case, Banshee obtains 12-16\% improvement (in green) on the rough estimation considering the instruction count only. For a large 32x32 \gls{mimo} input, the error of Banshee is only 0.6-20\%.

Despite the cycle performance over-estimation, in \Cref{fig:parmmse_cycles} we note that measuring the cycle count in Banshee does not invalidate the key insight on the speedup of \gls{simd} implementations, compared to \textit{16bHalf}. The values measured for 32x32 inputs, 1.05$\times$, 1.54$\times$, 1.07$\times$, confirm \textit{16bCDotp} as the fastest option, followed by \textit{8bwDotp} and \textit{16bwDotp}. The tool is useful for rough estimations of the \gls{E2E} cycle count. Targeting an open-source \gls{dut} facilitates further calibration and reduction of performance over-estimation based on \gls{rtl} simulations results.

\subsection{BER extraction via fast \gls{mc} simulation}

A classic telecommunications \gls{kpi} is the \gls{ber}, which measures the ratio of bit errors compared to the original transmitted bit sequence. We compute the \gls{ber} of the \gls{mmse} implemented on TeraPool simulating \gls{mc} iterations with Banshee and leveraging its bit-true functional modeling accuracy. For different input \gls{snr}, we iterate to a target error count. We explored the performance of the \gls{mmse} detector in different arithmetic precisions according to three variables: \gls{mimo} problem size, bit-sequence modulation, and wireless channel type. 

\begin{figure}[ht!]
    \centering
    \includegraphics[width=0.24\textwidth]{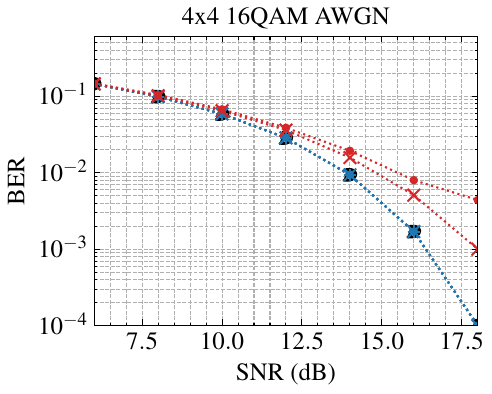}
    \includegraphics[width=0.24\textwidth]{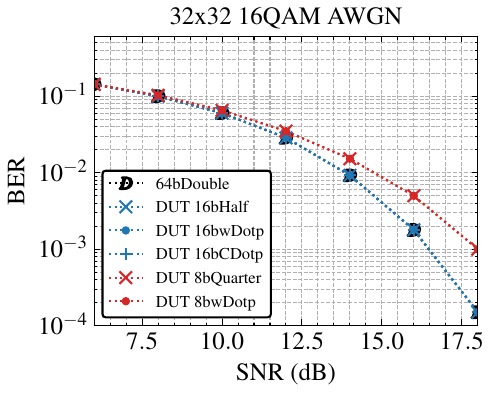}
    \includegraphics[width=0.24\textwidth]{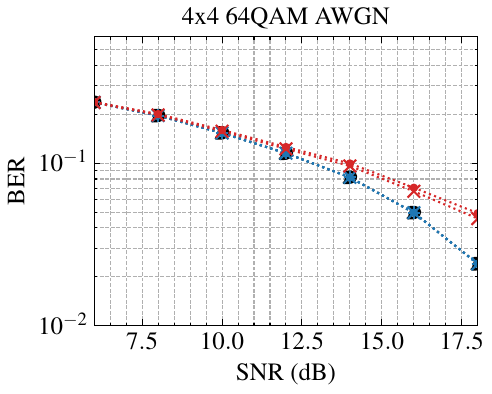}
    \includegraphics[width=0.24\textwidth]{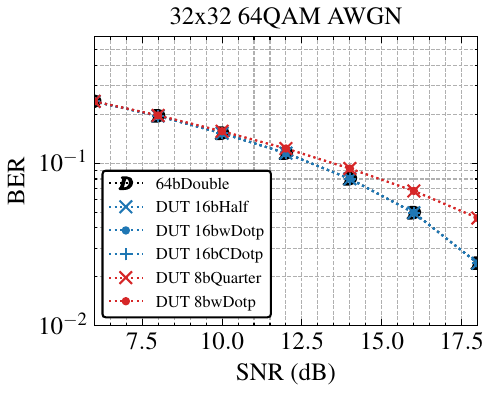}
    \caption{BER for 16\gls{qam} and 64\gls{qam} modulation, \gls{awgn} channel.}
    \label{fig:ber_16QAM_awgn}
    \vspace{-1em}
\end{figure}

The input bit-sequence maps to different \gls{qam} constellations. In \Cref{fig:ber_16QAM_awgn}, we compare the \gls{ber} vs. \gls{snr}, obtained for a 4x4 or a 32x32 \gls{mmse} problem with 16\gls{qam} and a 64\gls{qam} modulation. We observe the \textit{16bHalf}, \textit{16bwDotp}, and \textit{16bCDotp} implementations achieving the same results as the \textit{64bDouble} Python model. The \textit{8b} implementations both suffer the truncation of results precision before the \textit{16b} matrix inversion, corresponding to a 10$\times$ loss at 18dB. 

\begin{figure}[ht!]
    \centering 
    \includegraphics[width=0.24\textwidth]{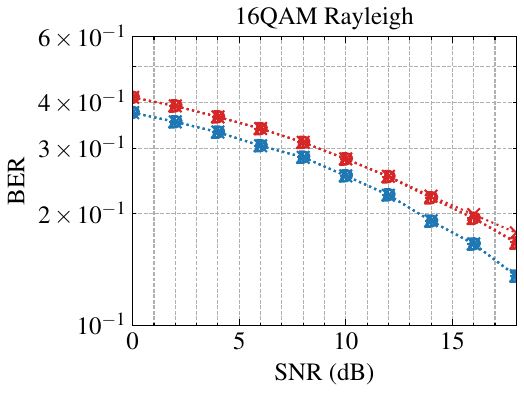}
    \includegraphics[width=0.24\textwidth]{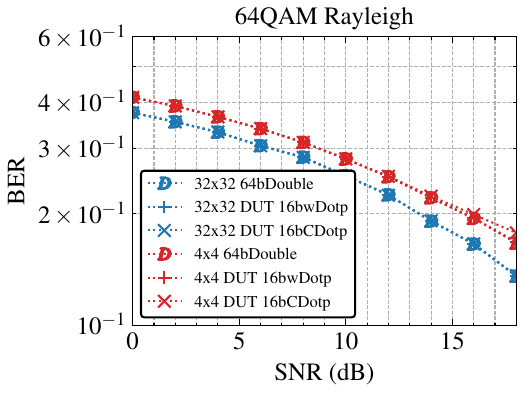}
    \caption{BER for 16\gls{qam} and 64\gls{qam} modulation, Rayleigh channel.}
    \label{fig:ber_rayleigh}
    \vspace{-1em}
\end{figure}

In \Cref{fig:ber_rayleigh}, we examine the \gls{ber} of the receiver, assuming a flat-fading Rayleigh channel. While the \gls{awgn} channel assumes zero attenuation and interference from other transmitters, the Rayleigh channel considers the effects of multi-path fading and models a \gls{5g}-\gls{mimo} transmission more closely. The \gls{mc} co-simulation of TeraPool and the wireless channel gives immediate feedback on the performance of low-bit arithmetic precision RISC-V extensions on the full wireless application: only the \textit{16bwDotp} and \textit{16bCDotp} implementations follow the performance of the \textit{64bDouble} golden model. The fast co-simulation approach revealed the benefits of accumulating in \textit{32b} (\textit{16bwDotp}), or execution of the complex \gls{mac} in \textit{32b} internal precision (\textit{16bCDotp}). 

\section{Conclusions}

This work presented an open-source \gls{sbt}-based lightweight simulator coupled to wireless channel models to extract the \gls{E2E} performance of RISC-V \gls{sdr}-transceivers. The simulation framework fulfills the requisites~\cite{Wittig_modeling5G_2020} to model \gls{bs} programmable receivers in large-scale \gls{ran} testbed.
\gls{sbt} offers fast, functional modeling of \gls{sdr} workloads on the open-source TeraPool 1024-core cluster, accounting for the low-bit precision of the hardware. It enables fast design space exploration of \gls{E2E} 5G transmissions, considering different modulation schemes and wireless channels: \gls{mmse} on a 5G \gls{ofdm}-symbol is simulated in only 9.5s-3min, depending on the input MIMO size, three orders of magnitude faster than \gls{rtl} simulation. Independent symbols are parallelized on up to 128 threads of a \gls{cots} server with 73-121$\times$ speedup compared to single-thread simulation.
The translation approach is agnostic of the algorithm used, which makes the method flexible to an evolving telecommunication standard. 
Finally, augmenting the simulator with a simple hardware timing model gives first-order estimates of the runtime on the \gls{dut}, with only 30\% average error, for different \gls{mimo} sizes. The fully open-sourced \gls{rtl} of the \gls{dut} ease further calibration and reduction of the performance over-estimation. 

\section*{Acknowledgment}
\textit{This work is funded in part by the COREnext project supported by the EU Horizon Europe research and innovation program under grant agreement No. 101092598.}

\bibliographystyle{IEEEtran} 
\bibliography{bibliography.bib}

\end{document}